\documentclass[12pt]{article}
\newcommand{\Comment}[1]{{}}
\usepackage{amsfonts,amsthm,amsmath,amssymb,slashed,mathrsfs}
\usepackage[textwidth = 430 pt, textheight = 630 pt]{geometry}
\usepackage{color}

\Comment{\usepackage{color}
\definecolor{MyDarkBlue}{rgb}{0.15,0.15,0.45}
\usepackage[linktocpage=true]{hyperref}
\hypersetup{
colorlinks=true,
citecolor=MyDarkBlue,
linkcolor=MyDarkBlue,
urlcolor=MyDarkBlue,
pdfauthor={Jeff Murugan and  Horatiu Nastase},
pdftitle={A 4D Duality Web},
pdfsubject={hep-th}
}

\usepackage[numbers,sort&compress]{natbib}
\usepackage{hypernat}}
\usepackage{graphicx}
\usepackage{cite}

\newcommand\ignore[1]{}
\def\one{{\,\hbox{1\kern-.8mm l}}}

\def\a{\alpha}\def\b{\beta}

\def\d{\partial}

\def\dslash{\partial\!\!\!/}

\newcommand{\Cset}{{\,\,{{{^{_{\pmb{\mid}}}}\kern-.45em{\mathrm C}}}}}

\newcommand{\be}{\begin{equation}}
\newcommand{\bea}{\begin{eqnarray}}

\newcommand{\ee}{\end{equation}}
\newcommand{\eea}{\end{eqnarray}}

\definecolor{orange-red}{rgb}{1.0, 0.27, 0.0}
\definecolor{darkviolet}{rgb}{0.58, 0.0, 0.83}

\begin{document}

\renewcommand{\thefootnote}{\fnsymbol{footnote}}

\makeatletter
\@addtoreset{equation}{section}
\makeatother
\renewcommand{\theequation}{\thesection.\arabic{equation}}

\rightline{}
\rightline{}


\vspace{10pt}


\begin{center}
{\huge \bf{A 4D Duality Web}}
\end{center} 
 \vspace{1truecm}
\thispagestyle{empty} \centerline{
{\Large {\bf{ Jeff Murugan}${}^{a}$}}\footnote{E-mail address: \Comment{\href{mailto:jeff.murugan@uct.ac.za}}{\tt jeff.murugan@uct.ac.za}}
{{ and}}
{\Large  {\bf{ Horatiu Nastase}${}^{b}$}}\footnote{E-mail address: \Comment{\href{mailto:horatiu.nastase@unesp.br}}
{\tt horatiu.nastase@unesp.br}}
                                                          }

\vspace{.5cm}

\centerline{{\it ${}^a$The Laboratory for Quantum Gravity \& Strings, }} 
\centerline{{\it Department of Mathematics and Applied Mathematics, }} 
\centerline{{\it University of Cape Town,}} 
\centerline{{\it Private Bag, Rondebosch, 7700, South Africa}}
\vspace{.3cm}
\centerline{{\it ${}^b$Instituto de F\'{i}sica Te\'{o}rica, UNESP-Universidade Estadual Paulista}} 
\centerline{{\it R. Dr. Bento T. Ferraz 271, Bl. II, Sao Paulo 01140-070, SP, Brazil}}

\vspace{1truecm}

\thispagestyle{empty}

\centerline{\sc Abstract}

\vspace{.4truecm}

\begin{center}
\begin{minipage}[c]{380pt}
{\noindent We construct a web of non-supersymmetric dualities in four spacetime dimensions for theories with theta 
and Maxwell terms. Our construction mirrors the recipe used in (2+1)-dimensions to demonstrate a similar duality for 
theories with Chern-Simons terms. As in the 3-dimensional case, the web consists of boson-boson, fermion-fermion 
and boson-fermion dualities, any one of which can be taken as a seed for the remaining two. To be concrete, we begin 
by constructing the basic bosonization duality which we then use to generate the web, using an analog of the 
3-dimensional massive construction given in \cite{Nastase:2017gxr} . The resulting 4-dimensional duality web is 
best understood as an extension of the, by now well-known, 3-dimensional case due to its somewhat singular 
nature. We conjecture that this is the final thread in a multi-dimensional duality web that connects low energy 
theories in 4-, 3- and 2-spacetime dimensions and speculate on its application to 3-dimensional topological insulators.
}
\end{minipage}
\end{center}

\vspace{.5cm}

\setcounter{page}{0}
\setcounter{tocdepth}{2}

\newpage

\renewcommand{\thefootnote}{\arabic{footnote}}
\setcounter{footnote}{0}

\linespread{1.1}
\parskip 4pt



\section{Introduction}

Duality, the idea of two entirely different descriptions of the same physical degrees of freedom,  is one of the most 
far-reaching and consequentual concepts in physics. Among the many examples that come to mind, we count:
\begin{itemize}
    \item {\it Electric-magnetic duality in 4-dimensional spacetime} - perhaps the first and most easily recognisable 
    example, the mapping $(\boldsymbol{E},\boldsymbol{B})\mapsto (\boldsymbol{B},-\boldsymbol{E})$ leaves 
    the vacuum Maxwell equations, 
   \begin{eqnarray}
      \boldsymbol{\nabla}\cdot \boldsymbol{E} &=& 0\,,\quad \boldsymbol{\nabla}\cdot \boldsymbol{B} = 0\,,
      \quad \boldsymbol{\nabla}\times \boldsymbol{E} = -\frac{\partial\boldsymbol{B}}{\partial t}\,,\quad 
      \boldsymbol{\nabla}\times \boldsymbol{B} = \frac{\partial\boldsymbol{E}}{\partial t}\,,
   \end{eqnarray}
   invariant. This seemingly innocuous observation, the precursor of S-duality, lies at the heart of Kapustin and 
   Witten's physical interpretation of the Geometric Langlands conjecture \cite{Kapustin:2006pk}.
   \item {\it Gauge/Gravity duality} - arguably, the most celebrated of the duality relations, the gauge theory/gravity 
   duality manifests itself most clearly in the AdS/CFT correspondence \cite{Maldacena:1997re} between a theory 
   of quantum gravity (such as superstring theory) in $AdS_{d}\times \mathscr{M}^{10-d}$ and a gauge theory on 
   the boundary of the $AdS$ component (such as supersymmetric Yang-Mills on $\partial$AdS$_{d}$).  Directly 
   or indirectly, it has been the primary catalyst for most of the advances both in gravity research as well as 
   strongly-coupled quantum field theory over the past two decades.
   \item {\it T-duality} - originally discovered in the context of the string worldsheet sigma model in string theory, 
   T-duality is a map that exchanges the geometric data associated with one 2-dimensional sigma model, 
   say\footnote{In this language $\mathscr{X}$ is a smooth manifold, equipped with a Riemannian metric 
   $g$ and Kalb-Ramond field $B$.} $(\mathscr{X},g, B)$ with another $(\widetilde{\mathscr{X}},\widetilde{g}, 
   \widetilde{B})$ such that the quantum field theories induced from such data are equivalent. For example, 
   the conformal field theory of a free boson on a circle of radius $R$ is equivalent to that of a free boson 
   on the circle of radius $1/R$ under T-duality. 
   \item {\it Bosonization} - is the remarkable observation, made more-or-less simultaneously in the condensed 
   matter and particle physics communities \cite{Mattis:1964wp,Luther:1975wr,Coleman:1974bu,Mandelstam:1975hb}, 
   that a theory of relativistic Dirac fermions with the usual anti-commutation relations, may be replaced by a bosonic 
   field theory that captures the same physics. This is particularly clean in one spatial dimension where, because of 
   the linear one-particle dispersion near the Fermi surface, particle-hole pairs exhibit a quasi-particle-like dispersion 
   near zero momentum and propagate as a coherent bosonic particle. 
\end{itemize}
Common to all of these examples is the fact that the duality maps one theory at strong (weak) coupling to another at 
weak (strong) coupling. Moreover, all but the gauge/gravity duality in the list above can be expressed formally in terms 
of a path integral. Roughly, one starts with a path integral, 
\begin{eqnarray}
    Z_{\mathrm{master}} = \int\mathscr{D}\Phi \mathscr{D}\Lambda\,\exp\left(S_{\mathrm{master}}[\Phi, \Lambda]\right)\,,
\end{eqnarray}    
over a space of auxillary fields and shows that integrating out one or other set of fields produces the path integral 
corresponding to the action function $S[\Phi]$ or its dual $\widetilde{S}[\Lambda]$. While this path integral formulation 
was well known for T-duality, it was not until the seminal work of Burgess and Quevedo \cite{Burgess:1993np} that it 
was understood that 1+1-dimensional bosonization could also be understood as a duality in this sense.  With a bose-fermi 
duality (bosonization) and a bose-bose duality (T-duality) in place, it seems like it was only a matter of time before the dots 
were connected to reveal a duality web in 2-dimensional spacetime.  Surprisingly, this was only carried out in the pedagogical 
work of Karch {\it et.al.} in 2019 \cite{Karch:2019lnn}, with the use of some sophisticated tools from the theory of quadratic 
forms and lessons learnt from condensed matter theory, in the intervening years.

\noindent
One such lesson was the realization that matter coupled to Chern-Simons gauge theory in 2+1-dimensions exhibits a 
remarkably rich structure. This is beautifully captured by Aharony's conjectured Chern-Simons-matter dualities\cite{Aharony:2015mjs},
\begin{eqnarray*}
   N_{f}\,\,\mathrm{fermions} +  U(k)_{-N+\frac{N_{f}}{2},-N+\frac{N_{f}}{2}} &\longleftrightarrow &
   N_{f}\,\,\mathrm{scalars} +  SU(N)_{k}\,,\\
   N_{f}\,\,\mathrm{fermions} +  SU(k)_{-N+\frac{N_{f}}{2}} &\longleftrightarrow &
   N_{f}\,\,\mathrm{scalars} +  U(N)_{k,k}\,,\\
   N_{f}\,\,\mathrm{fermions} +  U(k)_{-N+\frac{N_{f}}{2},-N - k +\frac{N_{f}}{2}} &\longleftrightarrow &
   N_{f}\,\,\mathrm{scalars} +  U(N)_{k,k+N}\,,
\end{eqnarray*}
where the `+' denotes a coupling between the matter field and a Chern-Simons field with a particular gauge group, and 
$U(N)_{k,l}\equiv \left(SU(N)_{k}\times U(1)_{Nl}\right)/{\mathbb{Z}_{N}}$. In particular, the choice $N_{f}=N = k = 1$ 
identifies two interesting cases; a scalar at the Wilson-Fisher fixed point is dual to a fermion coupled to an abelian 
Chern-Simons gauge field, and a free fermion is dual to a gauged Wilson-Fisher scalar. 

\noindent Another lesson came from 
the condensed matter physics of a (3+1)-dimensional topological insulator whose (2+1)-dimensional boundary exhibits 
dual descriptions in terms of (i) a massless Dirac fermion coupled to an external gauge field, $\mathscr{L} = 
i\overline{\Psi}D\!\!\!\!/_{A}\,\Psi$, and (ii) another Dirac fermion coupled to a gauge field, say $a$, which is itself 
coupled to the external gauge field\footnote{As pointed out in \cite{Seiberg:2016gmd}, this statement is not precisely 
true since the gauge-invariance (mod $2\pi$) of the $A da/4\pi$ term is incompatible with standard Dirac quantization. 
For a detailed discussion, and resolution, of this subtle issue, see \cite{Seiberg:2016gmd}.}, $\mathscr{L}_{\mathrm{dual}} 
= i\overline{\chi}D\!\!\!\!/_{a}\,\chi + \frac{1}{4\pi}A da$. This was recognised in \cite{Wang:2015qmt, Metlitski:2015eka} as 
a fermionic version of the more well-known bosonic particle-vortex duality \cite{Burgess:2000kj, Murugan:2014sfa} 
between the $O(2)$ Wilson-Fisher scalar $\mathscr{L}_{\mathrm{WF}} = |D_{B}\Phi|^{2} - |\Phi|^{4}$ and its gauged 
counterpart, $\mathscr{L}_{\mathrm{AH}} = |D_{b}\phi|^{2} - |\phi|^{4} + \frac{1}{2\pi}b\,dB$, the 3-dimensional abelian Higgs model. \\

\noindent
These three, seemingly disparate, threads were recognised to be part of an larger web of non-supersymmetric dualities 
in the summer of 2016 \cite{Karch:2016sxi,Murugan:2016zal,Seiberg:2016gmd}, precipitating a slew of results from both 
high energy and condensed matter communities that have led to a far deeper understanding of the phases of quantum 
matter. For an excellent and pedagogical account of the arguments leading to the 3 dimensional duality web, and many of the 
subsequent developments, we refer the interested reader to the excellent reviews \cite{Carl:2019bbf, Senthil:2018cru} 
and references therein. As so eloquently put in \cite{Karch:2019lnn}; dualities beget dualities. Much of this begetting is 
realised through path integral manipulations \cite{Burgess:1994tm} that relate the partition functions in the dual pair and, 
on the face of it, and up to the existence of certain topological invariants, none of these manipulations appear to depend 
in a crucial way on the number of dimensions in the problem, which begs the question: \\

\noindent
{\it If duality webs exist in 2- and 3-dimensional spacetimes, could one also exist in 4D?}\\

\noindent
Given the utility of the 3 dimensional duality web in driving developments in ultra quantum matter - mostly planar electronic matter 
residing on the boundary of bulk (3+1)-dimensional topological insulators, but also other novel quantum matter such as 
graphene - over the past few years, it is clear that a similar set of relations describing the quantum phases of bulk physics 
would be equally important. There are, however, some subtle issues that need unpacking first. 

\noindent 
The most pressing among 
these is that, both in 2- and 3-dimensions, the seed of the duality web, from which the rest of the duality relations could 
be derived, is bosonization. In $D=d+1$ spacetime dimensions, the duality approach to bosonization of \cite{Burgess:1994tm} 
maps a theory with a massive Dirac fermion to one for a Kalb-Ramond $(d-1)$-form $B$. This  in turn can be formulated 
as a scalar field theory with derivative interactions. However, while the bosonic theory retains several desirable properties 
such as gauge invariance under $B\to B + d\omega$ with $(d-2)$-form $\omega$; others, such as locality, are lost for 
$D\geq4$.  For example, in the large $m$ limit, and in the notation above,
\begin{eqnarray*}
   -\overline{\psi}\left(\partial\!\!\!/+m + ia\!\!\!/\right)\psi\quad\longleftrightarrow\quad 
   -\frac{1}{k_{D}}\Omega\frac{1}{\Box}\Omega + \Omega a\,,
\end{eqnarray*}
where $a$ is an external gauge field to which the fermion couples, $k_{D}$ is a $D$-dependent constant and the 
bosonic field $\Omega$ is the Hodge dual of the field strength for the Lagrange multiplier field $\Lambda$. Still, 
there are reasons to be optimistic. The most compelling was already pointed out in \cite{Seiberg:2016gmd} where 
the authors  provide solid, yet circumstantial, arguments for the embedding of the 3-dimensional duality web into a 
4-dimensional spacetime in which the various dualities (particle-vortex, bose-fermi and fermi-fermi) reside on a 2+1 
dimensional boundary, coupled through a complex coupling $\tau \equiv \frac{\theta}{2\pi} + \frac{2\pi i}{e^{2}}$ to a 
half-space bulk action for the gauge fields. Inspired by this argument, the recent construction of a 4-dimensional 
fermion-fermion duality in \cite{Palumbo:2019tmg}, and following the 3-dimensional construction in given in 
\cite{Nastase:2017gxr}, we attempt here to provide a direct construction of a 4-dimensional duality web.\\

\noindent
While the path integral manipulations we employ are of course not new, the central premise of this article 
namely the existence of a fully 4 dimensional duality web is, to the best of our knowledge, a new and novel addition 
to the existing 2- and 3-dimensional duality webs. For its pedagogical value, we show how our construction 
fits in with the arguments given by Seiberg {\it et.al} in \cite{Seiberg:2016gmd} and  perform the dimensional 
reduction to the 3-dimensional web. A key feature of our construction, which it inherits from that 
in \cite{Nastase:2017gxr}, is that the duality is performed away from criticality; we gap the theory with 
a non-zero mass, carry out the duality at the level of the path integral and then tune the mass (which
we will view as a regulator, of sorts) to zero.\\
 
\noindent  
The paper is organised as follows; section 2 reviews the 3-dimensional construction of the basic bosonization 
map given in \cite{Nastase:2017gxr}, paying attention to the details of how this is used to thread together a 
web of boson-fermion, boson-boson and fermion-fermion dualities. In section 3, we lift this to (3+1)-dimensions 
where (modulo the locality issue above) the 4-dimensional analog of bosonization, at least in its realization 
as a duality, is known. More tricky to define is the higher dimensional analog of the bosonic particle-vortex 
duality. Here also, we review the recent fermion-fermion duality of \cite{Palumbo:2019tmg}, pointing out 
some subtleties that make it not quite what we want for the duality web.
In section 4, we bring these ingredients together to construct, directly at the path integral level, the 
4-dimensional duality web. We conclude with a discussion of some implications for the condensed matter 
physics of bulk topological insulators.

\section{A review of the 3-dimensional web construction}

Before proceeding to the 4-dimensional case of interest to us, it will be useful to establish some conventions and 
familiarise ourselves with the path integral manipulations required by reviewing the construction of the 3 dimensional duality 
web. While our primary interest will be with 4-dimensional spacetime, much of our notation will carry over from  3 dimensions.
To begin, bosons will generally be denoted $\phi$, fermions by $\psi$, external (non-dynamical) gauge fields by 
capitalised letters like $A = A_{\mu}\,dx^{\mu}$ and emergent gauge fields (usually arising from gauging some 
internal symmetries) by lower case letters such as $a = a_{\mu}\,dx^{\mu}$. A massive scalar has action
\begin{eqnarray}
   S_{\mathrm{scalar}}[\phi;A] = \int\! d^{3}x\,\left(|D\phi|^{2} - m^{2}|\phi|^{2}\right)\,,
\end{eqnarray}
where the operator $D\equiv d - iA$ minimally couples the scalar to an external $U(1)$ gauge field. We will also 
have need to add a quartic operator $-\alpha|\phi|^{4}$ term to this, in which case we'll denote the action 
$S_{\mathrm{boson}}[\phi;A;\alpha]$. Similarly, the action for a gapped Dirac fermion,  coupled to an external 
$U(1)$ gauge field, will be given by
\begin{eqnarray}
   S_{\mathrm{fermion}}[\psi; A] = \int\! d^{3}x\,\,i\overline{\psi}\left(D\!\!\!\!/ + m\right)\psi\,,
\end{eqnarray}
with $D\!\!\!\!/ = \gamma^{\mu}D_{\mu}$ as usual. Central to the transmutation of statistics in 2+1-dimensions 
is the idea of flux attachment; adding a single flux quantum to a (fermion) boson changes it into a (boson) 
fermion. In a relativistic theory such as the ones considered in this section, this is accomplished by 
coupling to a Chern-Simons term\footnote{In this, and what follows, expressions like $A\,dA$ are 
shorthand for $A\wedge dA$ which, in turn is the differential form notation for the indexed expression 
$\epsilon^{\mu\nu\lambda}A_{\mu}\partial_{\nu}A_{\lambda}$.}
\begin{eqnarray}
   S_{\mathrm{CS}}[A] = \frac{k}{4\pi}\int A\,dA\,,
\end{eqnarray}
with Chern-Simons level $k\in\mathbb{Z}$, to ensure that this term is gauge invariant if $F=dA$ is canonically 
normalized. More relevant for our 4-dimensional construction is the related BF-coupling 
\begin{eqnarray}
   S_{\mathrm{BF}}[A;B] = \frac{1}{2\pi}\int A\,dB = S_{\mathrm{BF}}[B;A] + \mathrm{boundary\,\,term}\,,
\end{eqnarray}
with the coefficient chosen so that the flux $\displaystyle \frac{1}{2\pi}\int dB$ has unit charge under $A$. Having set up 
our notation, let us be clear about what we mean by `duality'. We will call two theories dual to each other in the infrared, 
if their partition functions are equal. Establishing this equality is non-trivial but, fortunately, at least algorithmic. 
Following \cite{Burgess:1994tm}, to dualise a given theory, say theory A, with action $S_{\mathrm{A}}[\phi]$, we:
\begin{itemize}
   \item Gauge an internal symmetry of the theory, which introduces a dynamical gauge field $a$ over which a 
   functional integral is to be performed.
   \item Constrain $a$ to be pure gauge by imposing the flatness condition
   $f = da = 0$ through a Lagrange multiplier $\Lambda$ in the path integral. This ensures that we are not adding 
   any additional degrees of freedom. Taken together, this produces a master action 
   $S_{\mathrm{master}}[\phi; a; \Lambda]$.
   \item Integrating out $\Lambda$, followed by $a$ results in the original theory A. On the other hand, reversing 
   the order and integrating out $a$, followed by $\phi$ produces a new theory, B say, with action 
   $S_{\mathrm{B}}[\Lambda]$,  for the dual field $\Lambda$.
\end{itemize} 
This entire process, including subtleties introduced into the path integral measure through gauge-fixing 
and a Fadeev-Popov determinant, is then summarised, rather tersely, as $S_{\mathrm{A}}[\phi] 
\longleftrightarrow S_{\mathrm B}[\Lambda]$. To illustrate this construction let us now review the 
3-dimensional duality web, following closely the form derived in \cite{Karch:2016sxi,Karch:2016aux,Nastase:2017gxr}. 
As stated above, the basic web consists of three interconnected dualities - 
bose-bose, fermi-fermi and bose-fermi - and while it is true that any of these can serve as a seed 
for the full web, we will usually begin with the last and work our way back.

\subsection{Bose-Fermi}

With this in mind, let us define the basic 3-dimensional bosonization duality. In its mass deformed version\footnote{Here the 
mass term for the fermion is to be thought of as a regulator taking one away from the quantum critical point at $m=0$. We 
shall have more to say about this in the 4-dimensional case later.}, as defined in \cite{Karch:2016aux}, it is a map between 
a massive Dirac fermion coupled to an external gauge field $A = A_\mu\,dx^{\mu}$ whose Chern-Simons flux we denote 
$\mathrm{CS}[A]$, and a complex scalar coupled to the same gauge field. At the level of the partition function, this is a 
statement of equality between the fermion partition function,
\begin{eqnarray}
   Z_{\rm fermion}[A;m]\equiv \!\!\int \!\! {\mathscr D}\psi \exp\Bigl(iS_{\mathrm{fermion}}[\psi;A]\Bigr)\;;
\end{eqnarray}
and that of the scalar with flux attached,
\begin{eqnarray}
   Z_{\rm scalar+flux}[A] &=&\!\!\!\int \!\!{\mathscr D}a{\mathscr D}\phi
  \exp\Bigl({iS_{\rm scalar }[\phi;a]+iS_{\rm CS}[a] +i S_{\rm BF}[a,A]}\Bigr)\,
\end{eqnarray}
via the relation
\be
Z_{\rm fermion}[A;m]e^{-\frac{i}{2}S_{\rm CS}[A]}= Z_{\rm scalar+flux}[A].\label{bosonization}
\ee
At this juncture, there are a few points that warrant mention. The first is that \eqref{bosonization} is, as alluded to earlier, 
really the result of implementing the full duality algorithm above. The second is that this equality of partition functions 
holds in the {\it infra-red} limit in which we send the quartic scalar coupling $\alpha\to\infty$ while simultaneously 
tuning the scalar mass to zero in order to hit the Wilson-Fisher fixed point. Writing the complex scalar as $\phi = \phi_{0}e^{i\theta}$, the relevant relation is then really 
\bea
 Z_{\rm scalar+flux}[A]&=&\!\!\!\!\!\!\!
 \lim_{{\!\!\a\rightarrow \infty,E\ll \a}}\!\int\!\!\!{\mathscr D}a_\mu {\mathscr D}\phi_0{\mathscr D}\theta {\mathscr D}\sigma\,\times\nonumber\\
 &\times&\!\!\! \!\exp\Biggl\{\!iS_{\rm scalar }[\theta,a;\phi_0]\!+\!iS_{\rm CS}[a]\!+i\!S_{\rm BF}[a,A]\nonumber\\
 &-i&\!\int\! d^3x\! \left[\frac{1}{2}(\d_\mu\phi_0)^2
 +\sigma(\phi_0^2-m)+\frac{\sigma^2}{2\a}\right]\Biggr\}.\nonumber\\
 &=&\int {\mathscr D}a_\mu {\mathscr D}\theta\,
 \exp\Bigl(iS_{\rm scalar }[\theta,a;\phi_0]+iS_{\rm CS}[a] +i S_{\rm BF}[a,A]\Bigr)\,,
 \label{equ}
\eea
with  $S_{\rm scalar}[\theta,a;\phi_0] \equiv -\frac{1}{2}\int d^3x\, \phi_0^2\,(\d_\mu\theta+a_\mu)^2$. Incidentally, this same limit suppresses 
the Maxwell term $(da)^{2}/4e^{2}$ for the dynamical gauge field relative to its Chern-Simons term since $e\to\infty$. 
Finally, to construct the duality web, we will need the time-reversed version of \eqref{bosonization}. Since, under the 
time-reversal operator, each of the BF and CS terms pick up a minus sign only, we find that
\be
   Z_{\rm fermion}[A]e^{+\frac{i}{2}S_{\rm CS}[A]}=\widetilde{Z}_{\rm scalar+flux}[A]\equiv 
   \!\!\int\!\! {\mathscr D}\phi {\mathscr D}a \exp\Bigl(iS_{\rm scalar}[\phi,a]-iS_{\rm CS}[a]
   -iS_{\rm BF}[a,A]\Bigr).   
   \label{timerev1}
\ee
Equations \eqref{bosonization} and \eqref{timerev1} can then be taken as seeds for the remaining 
fermi-fermi and bose-bose dualities as follows. At this point, they are postulated, but in the final subsection of this 
section we will show how they were proven, in a certain limit, in \cite{Nastase:2017gxr}

\subsection{Fermi-Fermi}

The fermi-fermi thread of the duality web builds on the seminal conjecture of Son that the composite fermion that so 
successfully explains a number of phenomena associated to the fractional quantum Hall effect is a Dirac fermion. 
In a nutshell, this duality posits that a massless Dirac fermion $\psi$, coupled to an external field $A$ is dual to a 
composite Dirac fermion $\chi$ coupled to a dynamical gauge field $a$ which is itself coupled to $A$ through a 
BF coupling. The latter theory is simply QED$_{3}$ with a single fermion flavour and an additional BF coupling 
between the external gauge field and a fluctuating one\footnote{Strictly speaking, one should call this ``BF-QED$_3$", but we will drop the ``BF" in what follows, for simplicity}. It is described by the partition function
\begin{eqnarray}
  Z_{\rm QED_{3}}[A;m]= \!\!\int\!\! {\mathscr D}\chi {\mathscr D}a \exp\Bigl(iS_{\rm fermion}[\chi,a] + \frac{i}{2}S_{\rm BF}[a,A]\Bigr)\,,   
\end{eqnarray}
so that Son's duality reads 
\be
Z_{\rm QED_{3}}[A]=Z_{\rm fermion}[A]\;.\label{Son}
\ee
This relation follows from the basic bosonization duality as follows: 
\begin{itemize}
\item Starting from the basic bosonization relation (\ref{bosonization}), we promote external gauge field 
$A$ to  a dynamical one $a$ with a BF-coupling $\frac{i}{2}S_{\rm BF}
[\bar a;A]$ to a new external field, also denoted by $A$ and then integrate over $a$ in the functional integral. 
\item With this, the left-hand side of \eqref{bosonization} becomes
$Z_{\rm QED_{3}}[A]$ as above. The right hand side on the other hand becomes
\begin{eqnarray}
   Z_{\rm scalar+fluxes}[A]&=&\!\!\int\!\! \mathscr{D}\phi \mathscr{D}a \mathscr{D}\widetilde{a}\,
\exp\Bigl(iS_{\rm scalar}[\phi;\widetilde{a}]+iS_{\rm CS}[\widetilde{a}]\cr
&+&iS_{\rm BF}[\widetilde{a},a] +\frac{i}{2}S_{\rm BF}[a,A]+\frac{i}{2}S_{\rm CS}[a]\Bigr).
\label{mama}
\end{eqnarray}

\item Now, integrating over the statistical field $a$, using its equation of motion, $da=-(dA+2d\widetilde{a})$, 
 substituting back into the action, and collecting terms gives  
\be
  e^{-\frac{i}{2}S_{\rm CS}[A]}\int\!\! \mathscr{D}\phi \mathscr{D}\widetilde{a}\, \exp\Bigl(iS_{\rm scalar}[\phi;\widetilde{a}] 
  - iS_{\rm CS}[\widetilde{a}]-iS_{\rm BF}[\widetilde{a},A] \Bigr)\;,
\ee
in which we recognise the functional integral as the scalar side of the time-reversed bosonization relation (\ref{timerev}). 
Together with the contact interaction $e^{+\frac{i}{2}S_{\rm CS}[A]}$ then, this equates finally to $Z_{\rm fermion}[A]$, 
establishing the fermionic particle-vortex duality. 
\end{itemize}

\subsection{Bose-Bose}

The final thread in the 3D duality web is the duality between two bosonic theories. The derivation of this duality 
hinges on an intermediate bosonization step that amounts to attaching flux to a fermion in the process transmuting 
it into a boson so it will be worth our while to clarify this first before moving on. To attach a background flux to a 
fermion, we start with the fermion partition function $Z_{\rm fermion}[A]$, promote the external gauge field to a 
dynamical one, $b$ say, then  couple the latter to another external gauge field $A$ through a BF coupling and, 
finally, integrate over $b$ in the functional integral. The result is that
\begin{eqnarray}
Z_{\rm fermion+flux}[A] = \!\!\int\!\! {\mathscr D}b\, Z_{\rm fermion}[b]e^{-\frac{i}{2}S_{\rm CS}[b]-iS_{\rm BF}[b,A]}\;.
\end{eqnarray}
On the other hand, the basic bosonization duality \eqref{bosonization}, replaces $Z_{\rm fermion}[b]$ with $Z_{\rm scalar 
+ flux}[b]e^{\frac{i}{2}S_{\rm CS}[b]}$ in the functional integral, and since the 1-form $b$ appears linearly, the integration 
over $b$ can be carried out using its equation of motion
$db=dA$ or, in the absence of any holonomies\footnote{We are glossing over some important, and technical points here 
since they are not relevant for our present discussion. \cite{Karch:2016sxi} is an absolute treasure-trove of insight on this 
and other subtlties.} substituting $b=A$ into the partition function to obtain $Z_{\rm scalar}[A]e^{iS_{\rm CS}[A]}$. 
In summary then, we have the new bosonization relation 
\begin{eqnarray}
   Z_{\rm fermion+flux}[A]=Z_{\rm scalar}[A]e^{iS_{\rm CS}[A]}\,,
   \label{newbosonization}
\end{eqnarray}
which also has a handy time-reversed version where, as usual, we change the sign of the CS and BF couplings to give
\begin{eqnarray}
   Z_{\rm scalar}[A]e^{-iS_{\rm CS}[A]}=\widetilde Z_{\rm fermion +flux}[A]\equiv \!\!\int\!\! {\mathscr D}a\, 
   Z_{\rm fermion}[a] \exp\Bigl(\frac{i}{2}S_{\rm CS}[a]+iS_{\rm BF}[a,A]\Bigr)\,.\nonumber\\
   \label{newtimerev}
\end{eqnarray}
Now we play exactly the same game with \eqref{newbosonization} in the form
\begin{eqnarray}
   Z_{\rm fermion+flux}[A]e^{-iS_{\rm CS}[A]}=Z_{\rm scalar}[A]\,.
\end{eqnarray}
On the left hand side, promoting the background $A$ to a dynamical $b$, coupling in a new background field, 
$A$ and functionally integrating results in 
\begin{eqnarray}
   \int\!\!{\mathscr D}a {\mathscr D}b\,\, Z_{\rm fermion}[b]\exp\Bigl( -\frac{i}{2}S_{\rm CS}[a]
   -iS_{\rm BF}[a;b]-iS_{\rm CS}[b] -iS_{\rm BF}[b,A]\Bigr) \,.
\end{eqnarray} 
Again, since $b$ enters this expression linearly, we can integrate it out via its equation of motion, $db = dA-da$. 
With the no-holonomies caveat in place we can replace $b$ with $A-a$ in the partition function to find (on the left-hand side)
\begin{eqnarray}
   \int\!\! {\mathscr D}a Z_{\rm fermion}[A]e^{\frac{i}{2}S_{\rm CS}[a]+iS_{\rm BF}[a,A]+iS_{\rm CS}[A]} = 
   \widetilde{Z}_{\rm fermion +flux}[A] e^{\frac{i}{2}S_{\rm CS}[A]} =  Z_{\rm scalar}[A]\;.
\end{eqnarray}
On the right hand side of \eqref{newbosonization}, the process of promoting the external gauge field to a dynamical one 
(and subsequent coupling to another external field) is another way of saying that we have {\it gauged} the global 
background $U(1)$. Consequently, the theory on the right hand side is simply scalar QED$_{3}$. Finally then, the 
boson-boson map reads
\be
Z_{\rm scalar-QED}[S]=Z_{\rm scalar}[S]\;.
\ee
If the scalar is massless with a quartic self-interaction, this is the statement that a Wilson-Fisher scalar is dual to a 
gauged Wilson-Fisher scalar, otherwise known as particle-vortex duality. 

\subsection{A constructive `proof' of the bosonization step}

To prove \cite{Nastase:2017gxr}
the bosonization step from the first subsection (and, by extension, the whole duality web, at least in the
low energy limit to be defined shortly),  one starts with the {\em particle-vortex duality} 
in the formulation of \cite{Murugan:2014sfa}, which is the duality between the {\it particle} partition function
\bea
  Z_{\rm particle}[\phi_0,A]\!\!&=&\!\!\int {\mathscr D}\theta\, e^{iS[\phi_0,\theta,A]}\cr
  &\equiv&\!\!\int\!\! {\mathscr D}\theta\, \exp\left[-i\!\!\int\!\! d^3x\, 
  \frac{1}{2}\left[(\d_\mu\phi_0)^2+\phi_0^2(\d_\mu\theta_{\rm smooth}+\d_\mu\theta_{\rm vortex} 
  +A_\mu)^2\right]\right]\cr
&&
\eea
and the {\it vortex} partition function 
\bea
   Z_{\rm vortex}[\phi_0,A]&=&\!\!\int\!\! {\mathscr D}\lambda_\mu\, e^{iS_{\rm dual}[\phi_0,\lambda,A]} \cr
  &=&\!\!\int\!\! {\mathscr D}\lambda_\mu \exp\Biggl[-i\!\!\int\!\! d^3x\Biggl(\frac{1}{2}(\d_\mu\phi_0)^2 + 
  \frac{1}{4(2\pi\phi_0)^2}\Lambda_{\mu\nu} \Lambda^{\mu\nu}\nonumber\\
  &+&\frac{1}{2\pi}\epsilon^{\mu\nu\rho}\lambda_\mu\d_\nu A_\rho+ j^\mu_{\rm vortex}\lambda_\mu\Biggr)\Biggr]\;,
\nonumber\\
\label{pvdual}
\eea
with $\Lambda_{\mu\nu}\equiv\partial_{[\mu}\lambda_{\nu]}$. If, in addition we consider $\phi_0$ constant and at low energies 
where $E\ll \phi_0^2$, we can drop the first two terms in the above. 

\noindent
Then we add $S_{\rm CS}[A]+S_{\rm BF}[A;C]$ to the actions in both particle and vortex path integrals (which adds 
``flux" to both sides), and then integrate over $A_\mu$ (formerly the electromagnetic field, now integrated over, so it 
is a ``statistical" gauge field) to obtain a function of $C_\mu$, which now plays the role of the electromagnetic field. On the 
``particle" side of the duality, we obtain the scalar+flux side of the basic bosonization step that we wanted to prove, 
\bea
  Z'_{\rm particle+flux}[C]&=&\!\!\int\!\!{\mathscr D}A_\mu{\mathscr D}\theta\, \exp\Bigl(iS_{\rm scalar}[\theta,A;  
  \phi_0]+iS_{\rm CS}[A]+iS_{\rm BF}[A;C]\Bigr)\cr
  &=&Z_{\rm scalar+flux}[C]\;,\label{nana}
\eea
while on the ``vortex" side we obtain 
\be
  Z'_{\rm vortex+flux}[C]=\!\!\int\!\! {\mathscr D}A_\mu{\mathscr D}\lambda_\mu\, \exp\left(iS_{\rm BF}.  
  [\lambda;A]+iS_{\rm BF}[A;C]
  +iS_{\rm CS}[A]+i\!\!\int\!\! d^3x\, j^\mu_{\rm vortex}\lambda_\mu\right)\,.
  \label{papap}
\ee
Since the action is quadratic in $A_{\mu}$, to carry out the integral over $A_\mu$ we can use 
its equation of motion, $dA=-(dC+d\lambda)$, leading to 
\be
  Z'_{\rm vortex+flux}[C]=\!\!\int\!\! {\mathscr D}\lambda_\mu\, \exp\left(-iS_{\rm CS}[C]-iS_{\rm CS}[\lambda]-  
  iS_{\rm BF}[C,\lambda] + i\!\!\int\!\! d^3x\, j^\mu_{\rm vortex}\lambda_\mu\right)\,.
  \label{zpvf}
\ee
After a redefinition of $\lambda_\mu \to \lambda_\mu=\sqrt{2}\tilde \lambda_\mu+ 
C_\mu\left(-1+\frac{1}{\sqrt{2}}\right)$, we finally obtain the gauge side of the BQ map, with a CS term for 
$C_{\mu}$ and a vortex term (which was not included in the original duality web),
\be
  Z'_{\rm vortex+flux}[C]=Z'_{\rm gauge+vortex}[C]\exp\left[-\frac{i}{2}S_{\rm CS}[S]-i\!\!\int\!\! 
  d^3x\, j^\mu_{\rm vortex}C_\mu
  \left(-1+\frac{1}{\sqrt{2}}\right)\right]\,.
  \label{tata}
\ee
The BQ map defined in \cite{Burgess:1993np,Burgess:1994tm}, is between the fermion partition function 
coupled to the field $C_\mu$ and a gauge action coupled to the same, 
\bea
  Z_{\rm fermion}[C;m]\!\!&=&\!\!Z_{\rm gauge}[C]=\!\!\int\!\! {\mathscr D}\lambda_\mu\, \exp\left(\frac{-i} 
  {2k_3}\epsilon^{\mu\nu\rho}
  \lambda_\mu \d_\nu\lambda_\rho-i\epsilon^{\mu\nu\rho}\lambda_\mu\d_\nu C_\rho\right)\cr
  &=&\!\!\int\!\! {\mathscr D}\tilde\lambda_\mu\, \exp\Bigl(-2iS_{\rm CS}[\tilde \lambda]
  -iS_{\rm BF}[\tilde \lambda,C]\Bigr)\,.
\eea
In order to find the gauge+vortex partition function above, we must replace $C_\mu\rightarrow 
C_\mu+\d_\mu\theta_{\rm vortex}$, giving
\bea
  & & Z_{\rm fermion+vortex}[C;m]= \!\!\int\!\! {\mathscr D}\psi {\mathscr D}\bar\psi\, 
  \exp\left[i\!\!\int\!\! \bar\psi\left(\dslash+m +C +\dslash 
  \theta_{\rm vortex}\right)\psi\right],\nonumber\\
  & & Z_{\rm gauge+vortex}[C] =\!\!\int\!\! {\mathscr D}\tilde\lambda_\mu\, \exp\left[-2iS_{\rm CS}[\tilde\lambda]
  -iC_{\rm BF}[\tilde \lambda,C] -i\!\!\int\!\! d^3 x\, j^{\mu}_{\rm vortex} \tilde\lambda_\mu\right],\qquad\\
  & & 
  Z_{\rm fermion+vortex}[C;m]= Z_{\rm gauge+vortex}[C] .\nonumber
  \label{zaza}
\eea
In summary then, we get the basic bosonization duality step we wanted plus a vortex component,
\bea
  Z_{\rm scalar+flux}[C]\!\!&=&\!\!Z'_{\rm particle+flux}[C]=Z'_{\rm vortex+flux}[C],\nonumber\\
 \!\!&=&\!\!Z'_{\rm gauge+vortex}[C]\exp\left[-\frac{i}{2}S_{\rm CS}[C]-i\!\!\int\!\! d^3x\, j^\mu_{\rm vortex}
 C_\mu\left(-1+\frac{1}{\sqrt{2}}\right)\right],\quad\qquad\\
  \!\!&=&\!\!Z'_{\rm fermion+vortex}[C]\exp\left[-\frac{i}{2}S_{\rm CS}[C]-i\!\!\int\!\! d^3x\, j^\mu_{\rm vortex}
  C_\mu\left(-1+\frac{1}{\sqrt{2}}\right)\right].\nonumber
  \label{finalresult}
\eea
We see that the basic bosonization step was proven constructively, at least in the low-energy limit $E\ll m=\phi_0^2$, $E\ll \a$, 
the same parameter as in \eqref{equ}. Of course, this does {\it not} imply a proof outside this limit (for $m\rightarrow 0$, 
in particular), which is what one really wants for the duality web. 

In four dimensions, we will follow this same construction of the basic bosonization step, in order to establish a duality web. 
We will find that it only holds in certain limits so there again, we do not claim to {\em prove} the 4-dimensional duality web, 
but rather to {\em define} it, constructively in the appropriate limit.

\section{Four-dimensional dualities}

So, the 3-dimensional duality web furnishes an interconnected set of relations between bosonic and fermionic theories. 
As we have reviewed in the previous section, there are three such classes of relations; bose-bose, fermi-fermi and bose-fermi, 
together with their time-reversed counterparts. From this construction, we have learnt a few interesting lessons; the first is 
that the Chern-Simons coupling, peculiar to odd-dimensional spacetimes, is a key player in this construction and the second 
is that, while any of the dualities can seed the whole web, it really is easiest to begin with a bosonization. Each of these 
is important to keep in mind for extending the web away from three dimensions. In this section we will follow the procedure 
laid out in the previous subsection in order to directly construct a basic 4-dimensional bosonization step analogous to the 
one in three dimensions, which will then be used to seed a 4 dimensional web.\\ 

\noindent
The ingredients for this construction are the 4-dimensional analogs of particle-vortex duality and the bose-fermi duality 
that we call the BQ map, described in \cite{Burgess:1994tm}. Fortunately for us, the latter was already constructed 
in  \cite{Burgess:1994tm}. The analog of particle-vortex duality is, however, a little more subtle. Particle-vortex duality 
(see, for example, \cite{Murugan:2014sfa}) in three dimensions is, of course, based on the Poincar\'{e} duality between 
a scalar and a 1-form gauge field. One dimension higher, a scalar is Poincar\'e dual to a 2-form ``gauge field" 
corresponding to the antisymmetric tensor $B_{\mu\nu}$.  Physically, this would point to a particle/vortex-string duality. 
An attempt at such a duality was considered in \cite{Beekman:2010zx}) in the context of disordered superfluids in 
higher dimensions where it was shown that a particle-vortex duality persists in any $D\geq 3$ with the role of the 
``vortex" played by codimension-2 branes. While certainly intriguing, for our purposes of constructing a 4-dimensional 
duality web, we construct a new version of this 4 dimensional particle-vortex-type duality inspired by the related construction 
in  \cite{Palumbo:2019tmg}. It will be instructive to review the latter before proceeding further. 

\subsection{A 4 dimensional fermi-fermi duality}

In an interesting recent article\cite{Palumbo:2019tmg}, the author argues for the equality between the actions for a Dirac fermion and a composite fermion in 4-dimensional spacetime. The argument starts from the action for a massive Dirac fermion coupled to a gauge field\footnote{In this section we retain the indices in various expressions to disambiguate between contractions with the metric and those involving the completely anti-symmetric tensor $\epsilon_{\mu\nu\lambda\sigma}$.}, 
\begin{eqnarray}
   S_1[\psi,A_\mu]=\int d^4x \Bigl[\bar\psi\Gamma^\mu(\d_\mu+iA_\mu)\psi+m\bar\psi\psi\Bigr]\;.   
   \label{fermi1}
\end{eqnarray}
Integration of the massive fermion and defining, as usual, $F_{\mu\nu} = \partial_{[\mu}A_{\nu]}$ results in a low-energy effective action
\begin{eqnarray}
   S_1^{\rm eff}[A_\mu]=\frac{1}{32\pi}\int\! d^4x\, \epsilon^{\mu\nu\rho\sigma}F_{\mu\nu}F_{\rho \sigma}\;,
   \label{bose1}
\end{eqnarray}
of the $\theta F\widetilde{F}$ form. Here, the mass $m<0$ of the fermion  is to be thought of as a regulator 
to be removed at the end of the computation. Similarly, the action 
\begin{eqnarray}
   S_2^{\rm eff}[A_\mu, a_\mu,B_{\mu\nu}] = \!\!\int\! d^4x\, \epsilon^{\mu\nu\rho\sigma}\left[\frac{1}{32\pi}  
   F_{\mu\nu}f_{\rho\sigma} -\frac{1}{8\pi}B_{\mu\nu}\left(F_{\rho\sigma}+f_{\rho\sigma})\right) 
   +\frac{1}{32\pi^2\chi}H_{\mu\nu\rho}H^{\mu\nu\rho}\right],\cr
\label{bose2}
\end{eqnarray}
with $f_{\mu\nu}\equiv\d_\mu a_\nu -\d_\nu a_\mu$, $H_{\mu\nu\rho}\equiv\d_\mu B_{\nu\rho}+\d_\nu B_{\rho\mu}
+\d_\rho B_{\mu\nu}$, and where $\chi$ is a real parameter, can be thought of as the low-energy effective action, 
obtained by integrating out the massive fermions in the action for a composite neutral Dirac fermion, $\Psi$ coupled 
to an  emergent gauge field $a_\mu$ and a new antisymmetric tensor gauge field $B_{\mu\nu}$, 
\bea
S_2[\Psi,A_\mu,a_\mu,B_{\mu\nu}]&=&\!\!
\int d^4x \left[\bar\Psi\Gamma^\mu(\d_\mu+ia_\mu)\Psi+m\bar\Psi\Psi\right.\cr
&&\left.-\frac{1}{8\pi}\epsilon^{\mu\nu\rho\sigma}B_{\mu\nu}(F_{\rho\sigma}+f_{\rho\sigma})+\frac{1}{32\pi^2\chi}H_{\mu\nu\rho}
H^{\mu\nu\rho}\right]\,.\label{fermi2}
\eea
Here again the fermion mass, $m<0$ is interpreted as a regulator to be removed at the end of the calculation. We note 
that, from the first two terms in $S_2^{\rm eff}$, $B_{\mu\nu}$ has dimension 2, which means that the last term is a 
dimension 6 operator, with $\chi$ a dimension 2 parameter. The upshot of this is that {\em at low energies} we can 
ignore the $H^2$ term in the effective action. Consequently, $B_{\mu\nu}$ acts as a Lagrange multiplier that 
enforces the constraint $ F_{\mu\nu}=-f_{\mu\nu}$, or since the gauge fields are Abelian and there are no 
nontrivial topological issues lurking, $A_\mu=-a_\mu$.
The implication then is that $S_2^{\rm eff}$ and $S_1^{\rm eff}$ are identical. It is then concluded in \cite{Palumbo:2019tmg}, 
correctly, that the Dirac fermion and composite fermion actions (\ref{fermi1}) and (\ref{fermi2}) are equal. Palumbo goes 
further and shows that, at the 3-dimensional boundary of the 4-dimensional spacetime, 
the effective action
\begin{eqnarray}
   S_1^{\rm eff}[A_{\mu}]=\int_{M_4}d{\cal L}_{\rm CS}=S_{\rm CS}[A]=\frac{1}{8\pi}\int_{\d M_4}\!\!\! 
   A\,dA\;,
\end{eqnarray}
while the effective action
\begin{eqnarray}
   S_2^{\rm eff}[A_\mu,a_\mu, B_{\mu\nu}]&=&\!\!\int_{M_4}d{\cal L}_{CS-BF}
   =S_{\rm CS-BF}[A,a,b] = \frac{1}{8\pi}\int_{\d M_4} \!\!\Bigl(a\,da - b\,d\left(a+A\right)\Bigr)\,,\cr
   &&
\end{eqnarray}
which is of the mixed CS-BF form.\footnote{To put the action in this form, we can either first peel off the derivative acting on 
$a +A$, then in the 3-dimensional action define the 2-form field strength $B \equiv db$, and partially integrate to put $S_{\rm 
CS-BF}$ in the above form, or equivalently, first write $B$ in terms of $b$ and peel off the derivative acting on $b$ to directly 
obtain the action.} He then notes that the 3-dimensional Chern-Simons action $S_{\rm CS}[A]$ is the effective action for a 
Dirac fermion $\psi$ coupled to $A$, the trivial 3 dimensional version of $S_1$, 
while $S_{\rm CS-BF}[A,a,b]$ is the effective action for a composite neutral Dirac fermion coupled to a (dynamical) 
emergent gauge field $a$, plus BF terms, 
\begin{eqnarray}
S_2[\Psi,A,a,b]=\int  
\left[\bar\Psi D\!\!\!\!/_{a}\Psi+m\bar\Psi\Psi-\frac{1}{8\pi}b\, d(a +A)\right].
\end{eqnarray}
All of this analysis is certainly correct but, while the similarities are certainly there, it is important to note that this
 is {\it not} the same as the duality in Son's conjecture. There, in the composite fermion action, the BF term 
 couples the dynamical gauge field $a$ to the external field $A$ via the term $+\frac{1}{4\pi} A\,da$, and not 
 $b$ to $a$ and $A$ as in the above. Neverthess, the idea proposed in \cite{Palumbo:2019tmg} is quite neat 
 and we take inspiration from it to propose an alternative. As we will show below, a similar 4-dimensional duality 
 can be realised by forgetting about the fermions, concentrating instead  on the {\em low energy} bosonic 
 duality between (\ref{bose1}) and (\ref{bose2}), and embedding it into another, 4-dimensional analog of the 
3-dimensional particle-vortex duality combined with a 4-dimensional version of the BQ map \cite{Burgess:1994tm}. 

\subsection{Bosonization in 4 dimensions}
In order to understand how to lift 3-dimensional particle-vortex duality to four dimensions, note that the former is a map between the Lagrangian
\begin{eqnarray}
-\frac{(\d_\mu\phi_0)^2}{2}-\frac{\phi_0^2}{2}(\d_\mu\theta_{\rm vortex}+\d_\mu\theta_{\rm smooth}+A_\mu)^2\,,
\end{eqnarray}
and
\begin{eqnarray}
-\frac{(\d_\mu\phi_0)^2}{2}-\frac{H_{\mu\nu}^2}{4\phi_0^2}+\epsilon^{\mu\nu\rho}b_\mu\d_\nu(a_\rho+A_\rho)\,.
\end{eqnarray}
Here we have written the complex scalar $\phi(x) = \phi_{0}(x)e^{i\theta(x)}$ and split the phase $\theta$ into a ``smooth" part $\theta_{\rm smooth}$ and a part, $\theta_{\rm vortex}$, that encodes the non-trivial monodromy of the vortex. $b_{\mu}$ is a Lagrange multiplier that comes from gauging the global $U(1)$ symmetry of the first Lagrangian, and $H_{\mu\nu}$ is its associated field strength. Finally, the (fluctuating) statistical gauge field $a_{\mu}\equiv \partial_{\mu}\theta_{\rm vortex}$. We have also dropped the Maxwell term for the external gauge field $A_{\mu}$ in the low energy limit where the duality is valid. This form of the duality extends in a natural way to a 4-dimensional low energy duality between
\begin{eqnarray}
  -\frac{(\d_\mu \phi_0)^2}{2}-\frac{H_{\mu\nu\rho}^2}{4\phi_0^2}-\frac{1}{8\pi }   
  \epsilon^{\mu\nu\rho\sigma} B_{\mu\nu}\d_\rho
   (a_\sigma+A_\sigma)+\frac{1}{32\pi}\epsilon^{\mu\nu\rho\sigma}f_{\mu\nu}f_{\rho\sigma}
\end{eqnarray}
and 
\be
-\frac{(\d_\mu\phi_0)^2}{2}-\frac{\phi_0^2}{2}(\d_\mu\theta+A_\mu)^2+\frac{1}{32\pi}
\epsilon^{\mu\nu\rho\sigma}F_{\mu\nu}F_{\rho\sigma}.
\ee
Here again,  we use the low energy condition to ignore the kinetic term for $B_{\mu\nu}$, so this is consistent 
with the 3-dimensional case. The only extra ingredient is the existence of the $\theta$-term. As it stands, this 
is still a boson-boson duality. Fortunately, the 4-dimensional Burgess-Quevedo (BQ) map in \cite{Burgess:1994tm} 
provides a relation between a fermion coupled to an external $U(1)$ gauge field and a 2-form field $B_{\mu\nu}$ 
(which is dual to a pseudoscalar in four dimensions) coupled to the same gauge field so, by composing these 
two duality relations, {\em like we saw that we did when constructively proving the bosonization step in 3 dimensions}, 
we should get the fundamental bosonization step for the 4-dimensional duality web.\\

\noindent
Before proceeding further, let's pause to make a comment. After the dust has settled, we will 
see that the combined BQ + PV map takes a Dirac fermion to a complex scalar (modulo
some topological fluxes), essentially because the BQ duality maps a Dirac fermion to a scalar, dual to the gauge 
field. In three dimensions, as far as degrees of freedom are concerned, this makes sense since a complex scalar has the 
same number of on-shell degrees of freedom (1 complex, or 2 real) as a Dirac fermion (2 complex components, reduced on-shell 
to 1). In four dimensions, however, the duality still maps a Dirac fermion to a complex scalar, only now, 
with the Dirac fermion carrying 2 complex degrees of freedom and the scalar carrying 1 complex degree 
of freedom, there appears to be a mismatch of on-shell degrees of freedom. How then do we make 
sense of the duality? It refers to the response to 
the outside coupling (electromagnetism, with source $A_\mu$), as understood by Burgess and Quevedo.
That is to say, we understand it as the equality of the partition functions for the systems coupled to an electromagnetic 
source $A_\mu$. This was precisely the case in the 3-dimensional web, only there the 
number of degrees of freedom of the systems coupled to external electromagnetic source were the same. 
The latter is nice to have, but not necessary. 

\noindent
In more detail, the particle/vortex-string duality in 4-dimensional spacetime is an equality between the partition functions
for ``particle" variables,
\begin{eqnarray}
  Z_{\rm particle}[A]&=&\int {\mathscr D}\theta\exp\left\{i\int d^4x\left[-\frac{(\d_\mu\phi_0)^2}{2}
  -\frac{\phi_0^2}{2}  (\d_\mu\theta+A_\mu)^2\right.\right.\cr
  &&\left.\left.+\frac{1}{32\pi}
  \epsilon^{\mu\nu\rho\sigma}F_{\mu\nu}F_{\rho\sigma}\right]\right\}\;,
\end{eqnarray}
and that for ``vortex" variables, 
\begin{eqnarray}
Z_{\rm vortex}[A]&=&\!\!\int\! {\mathscr D}a_\mu\, {\mathscr D}B_{\mu\nu}\,\exp\left\{i\int 
d^4x\left[-\frac{(\d_\mu \phi_0)^2}{2}-\frac{H_{\mu\nu\rho}^2}{4\phi_0^2}\right.\right.\cr
&&\left.\left.-\frac{1}{8\pi}\epsilon^{\mu\nu\rho\sigma} B_{\mu\nu}\d_\rho
(a_\sigma+A_\sigma) 
+\frac{1}{32\pi}\epsilon^{\mu\nu\rho\sigma}f_{\mu\nu}f_{\rho\sigma}\right]\right\}\,.
\end{eqnarray}
When $\phi_0$ is constant, and at energies $E\ll \phi_0$, we can drop the two first terms, just as in three dimensions. 
Then, imitating what we did in three dimensions, we add some terms like $S_{CS}[A]$ and $S_{BF}[A,C]$ 
and integrate over $A_\mu$ to get a function of $C_\mu$ only.  On the vortex side, modulo some subtlties, this should
furnish the gauge side of the BQ map, while the particle side would define the scalar part of the basic bosonization 
step of the duality web. The $A_\mu$ integration produces, among others, an $\epsilon^{\mu\nu\rho\sigma}B_{\mu\nu}G_{\rho\sigma}$ 
term, with $G_{\mu\nu} = \partial_{[\mu}C_{\nu]}$ . More generally though, in 4-dimensional spacetime, 
consider the situation when the Maxwell terms will 
dominate over the topological terms $\epsilon FF$ and $\epsilon FG$, so the latter can be ignored in most situations, i.e.,
\begin{eqnarray}
\frac{1}{g_4^2}\Bigl(F_{\mu\nu}F^{\mu\nu}+F_{\mu\nu}G^{\mu\nu}\Bigr) +\frac{1}{32\pi}\epsilon^{\mu\nu\rho\sigma}\Bigl(F_{\mu\nu}
F_{\rho\sigma}+F_{\mu\nu}G_{\rho\sigma}\Bigr)\,,
\end{eqnarray}
with $1/g_4^2\gg \theta$. Absorbing the $g_4$ into $A_\mu$ and $C_\mu$, and 
the $1/(4\pi)$ into $B_{\mu\nu}$ and carrying out the path integral over $A_\mu$ in the partition function, we find 
\begin{eqnarray}
   F_{\mu\nu}=\frac{\epsilon_{\mu\nu\rho\sigma}B^{\rho\sigma}-G_{\mu\nu}}{2}\;,
\end{eqnarray}
which, when replaced in the action gives 
\be
-B_{\mu\nu}B^{\mu\nu}+\frac{1}{64\pi}\epsilon^{\mu\nu\rho\sigma}B_{\mu\nu}G_{\rho\sigma}
-\frac{1}{4}G_{\mu\nu}G^{\mu\nu}.
\label{BBact}
\ee
However, it is a well-known fact that, unlike the local CS and BF terms in three dimensions, in $D\geq4$, 
the BQ map \cite{Burgess:1994tm}  produces a non-local $\Box^{-1}$ interaction. Specifically,  
\begin{eqnarray}
Z_{\rm fermion}[C,m]\!\!&=&\!\!Z_{\rm gauge}[C]\cr
&=&\!\!\int\! {\mathscr D}B_{\mu\nu}\,\exp\left(-i\!\!\int\!\! d^4x \left[\frac{1}{2k_4}\epsilon^{\mu\nu\rho\sigma}
\d_\nu B_{\rho\sigma}\frac{1}{\Box}\epsilon^{\mu\nu'\rho'\sigma'}\d_{\nu'}B_{\rho'\sigma'}
\right.\right.\cr
&&\left.\left.+\epsilon^{\mu\nu\rho\sigma}C_\mu\d_{\nu}B_{\rho\sigma}
\right]\right)\cr
&=&\!\!\int\!\! {\mathscr D}B_{\mu\nu}\,\exp\left(-i\!\!\int\!\! d^4x \left[\frac{1}{3k_4}H_{\mu\nu\rho}
\frac{1}{\Box}H^{\mu\nu\rho}+\frac{1}{2}\epsilon^{\mu\nu\rho\sigma} B_{\mu\nu}G_{\rho\sigma}\right]\right\}.\cr
&&
\end{eqnarray}
While this expression doesn't look much like what we had above, it is in fact equivalent. To see this, let 
$\Omega^\mu=\epsilon^{\mu\nu\rho\sigma}
\d_\nu B_{\rho\sigma}$. Then, the path integration over $B_{\mu\nu}$ produces the $\Omega$-equation of motion,
\begin{eqnarray}
   \Omega_\mu=-k_4\Box C_\mu\;.
\end{eqnarray}
When replaced back into the path integral, this results in a 
$+\frac{1}{2}k_4 C_\mu \Box C^\mu$ term.
On the other hand, the same variation with respect to $B_{\mu\nu}$ in (\ref{BBact}) gives 
$B_{\mu\nu}=+\frac{1}{4}\epsilon_{\mu\nu\rho\sigma}G_{\rho\sigma}$
and a corresponding contribution  of $+\frac{1}{4}G_{\mu\nu}G^{\mu\nu}$
in the path integral. In Lorenz gauge $\d^\mu C_\mu=0$, this then reduces to the same action as above. 
To summarise, we have obtained an action {\em equivalent} to that in the gauge side of the BQ map in four dimensions. 
Note that the action in $Z_{\rm gauge}[S]$ in the BQ map is 4-dimensional, but it is easy to 
imagine adding a 3-dimensional boundary term of the form $\epsilon^{\mu\nu\rho} b_\mu \d_\nu b_\rho$, 
which can be neglected in the bulk, but not on the boundary, and where $b_\mu$ is path 
integrated in three dimensions. This would then facilitate a dimensional reduction to three dimensions.\\

\noindent
Now, connecting the pieces of this argument; the scalar part of this basic bosonization step 
is given by the particle plus flux side of the particle/vortex-string duality, 
\bea
&&Z_{\rm scalar+flux}\equiv Z'_{\rm particle+flux}[C]
=\!\!\int\!\! {\mathscr D}A_\mu\, {\mathscr D}\theta\,\exp\left\{i\!\!\int\!\! d^4 x 
\left[-\frac{(\d_\mu\phi_0)^2}{2}-\frac{\phi_0^2}{2}(\d_\mu \theta+A_\mu)^2\right.\right.\cr
&&\left.\left. +\frac{1}{32\pi}\epsilon^{\mu\nu\rho\sigma}[F_{\mu\nu}F_{\rho\sigma}+F_{\mu\nu}G_{\rho\sigma}]
+\frac{1}{g_4^2}[F_{\mu\nu}F^{\mu\nu}+F_{\mu\nu}G^{\mu\nu}]\right]\right\}\;,\cr
&&
\eea
where we have added an $\epsilon FG$ term that matches the $\epsilon FF$ already present 
(and which, as we point out above, is subleading to the Maxwell term at small $g_4$ in 
four dimensions). Note that the first two terms in the action correspond to a complex 
scalar coupled to $A_\mu$ at the fixed point, as in the 3-dimensional story. The fermion side of the duality starts with the vortex + flux side, 
\bea
&&Z'_{\rm vortex+flux}[C]\cr
&=&\!\!\int\!\! {\mathscr D}A_\mu\, {\mathscr D}B_{\mu\nu}\,{\mathscr D}a_\mu\,
\exp\left\{i\int d^4x\left[-\frac{(\d_\mu \phi_0)^2}{2} -\frac{H^2_{\mu\nu\rho}(B)}{
4\phi_0^2}-\frac{1}{8\pi}\epsilon^{\mu\nu\rho\sigma}B_{\mu\nu}\d_\rho (a_\sigma+A_\sigma)\right.\right.\cr
&&\left.\left. +\frac{1}{32\pi}\epsilon^{\mu\nu\rho\sigma}(f_{\mu\nu}f_{\rho\sigma}+f_{\mu\nu}G_{\rho\sigma})
+\frac{1}{g_4^2}(F_{\mu\nu}F^{\mu\nu}+F_{\mu\nu}G^{\mu\nu})\right]\right\}\;,\cr
&&
\eea
where the first two terms are neglected in matching with the gauge side of the BQ map, and where we have converted $\epsilon FG$ 
from the particle side into a $\epsilon f G$ term on the vortex side, since at low energies $a_\mu=-A_\mu$.  
We note that the 3-dimensional vortex current, $j^\mu _{\rm vortex }C_\mu$ which came from 
$j^\mu_{\rm vortex} \lambda_\mu$ corresponds to the term $\epsilon^{\mu\nu\rho\sigma} B_{\mu\nu} 
\d_\rho \d_\sigma \theta_{\rm vortex}$ in four dimensions, 
with $\d_\sigma \theta_{\rm vortex}$ standing in 
for the statistical gauge field $a_\sigma$. Also note that for a vortex (or, more precisely in four dimensions, 
a vortex string)  $\d_\mu\theta_{\rm vortex} +A_\mu=0$. Finally, carrying out the $A_\mu$ and $B_{\mu\nu}$ 
path integrals (and neglecting the $\epsilon B\d a$ term as small relative to the Maxwell terms) 
leads to the fermion partition function, 
\be
Z_{\rm fermion}=\!\!\int\!\! {\mathscr D}a_\mu\,\left\{Z_{\rm fermion}[C;m] \times
\exp \left[+\frac{1}{32\pi} \epsilon^{\mu\nu\rho\sigma} (f_{\mu\nu}f_{\rho\sigma} -f_{\mu\nu}G_{\rho\sigma})\right]\right\}\;,
\ee
precisely the fermion side of the basic bosonization step, as anticipated. 


\subsection{Dimensional reduction to 3 dimensions}

Let us check now how these statements translate into three dimensions. First, we would like to understand 
how the $\epsilon B \d A$ term in four dimensions (in the vortex side of the particle-vortex duality) 
reduces to the correct $\epsilon b \d A$ term in three dimensions. To state the question more precisely, 
suppose that the boundary is in the $x^4$ direction in the 4-dimensional spacetime and take (with $i,j=1,2,3$)
\be
B_{4 i}=\d_4 b_i\;,\;\; B_{ij}=0\;,\;\; a_4=A_4=0.
\ee
With this ansatz, $\epsilon^{\mu\nu\rho\sigma} \d_\mu b_\nu \d_\rho (a_\sigma+A_\sigma)=
\epsilon^{4ijk}\d_4 [b_i \d_j(A_k+a_k)]$, while the field strength for $B_{\mu\nu}$ satisfies
\begin{eqnarray}
H_{i 4 j}=\d_i B_{4j}+\d_4 B_{ji}+\d_j B_{i4}=\d_4(\d_i b_j-\d_j b_i)\equiv\d_4 k_{ij}\;,
\end{eqnarray}
with all remaining $H_{ijk}=0$. Moreover, since 
\begin{eqnarray}
\frac{H_{\mu\nu\rho}H^{\mu\nu\rho}}{\chi}=\frac{\d_4 k_{ij}\d_4 k_{ij}}{\chi}\;,
\end{eqnarray}
this term as still be safely neglected. This leaves just the nonlocal $H_{\mu\nu\rho}\frac{1}{\Box}H_{\mu\nu\rho}$ term.
At low energies in three dimensions (and with an obvious definition of $\Box_{(D)}$), 
\begin{eqnarray}
\Box_{(4)} H_{i 4 j}=(\d_4^2+\Box_{(3)})H_{i4j}=(\d_4^2-l^2)H_{i4j}\simeq \d_4^2 H_{i4j}\,,
\end{eqnarray}
where small energies means $k^2\ll \d_4^2$. From this last expression then we deduce that in this limit, $1/\Box_{(4)}\simeq 1/\d_4^2$, so 
\begin{eqnarray}
H_{\mu\nu\rho}\frac{1}{\Box_{(4)}}H_{\mu\nu\rho}=H_{i4j}\frac{1}{\Box_{(4)}}H_{i4j}\simeq \d_4 k_{ij}\frac{1}{\d_4^2}\d_4k_{ij} 
=-k_{ij}k^{ij}\;,
\end{eqnarray}
after an integration by parts.This term is clearly negligible in three dimensions. Indeed, integrating over the 
(finite) 4-dimensional bulk gives $\int dx_4\equiv \frac{1}{\phi_0^2}$, one of the terms that was neglected in the 
3-dimensional calculation. Of course, the correct term to be added in the gauge side of the BQ map in three dimensions 
is $CS[b]$, which has to be added by hand as a boundary term in the 4-dimensional action. 
We have checked that the remaining terms in the action dimensionally reduces correctly, so that 
the basic 4-dimensional bosonization map reduces to the appropriate map in three dimensions.

\section{Constructing the duality web}

With all the pieces in place, we can now construct a 4-dimensional web of dualities, by a combination of the 
basic bosonization duality and its time-reversed version. In this section, we will demonstrate this by constructing
 the fermi-fermi duality and bose-bose duality through repeated application of the basic bosonization duality, 
 imitating closely the 3-dimensional version in  \cite{Karch:2016sxi}.

\subsection{A 4 dimensional fermi-fermi duality}

In order to imitate the 3-dimensional procedure in 4 dimensions, we need one more ingredient; the time-reversed 
version of the bosonization step. Reversing the time direction, gives the $\epsilon FF$ and $\epsilon FG$ terms a 
minus sign while leaving all the rest unchanged, so that the time-reversed
version of bosonization is
\bea
&&\!\!\int\!\! {\mathscr D} a_\mu \left\{Z_{\rm fermion}[C;m] \times
\exp \left[-\frac{1}{32\pi} \epsilon^{\mu\nu\rho\sigma} \Bigl(f_{\mu\nu}f_{\rho\sigma} -f_{\mu\nu}G_{\rho\sigma}\Bigr)\right]\right\}\cr
&=&\widetilde{Z}_{\rm scalar+flux}\,[C]
=\!\!\int\!\! {\mathscr D}A_\mu\, {\mathscr D}\theta\,\exp\left\{i\!\!\int\!\! d^4 x \left[-\frac{(\d_\mu\phi_0)^2}{2}
-\frac{\phi_0^2}{2}(\d_\mu \theta+A_\mu)^2\right.\right.\cr
&&\left.\left. -\frac{1}{32\pi}\epsilon^{\mu\nu\rho\sigma}\Bigl(F_{\mu\nu}F_{\rho\sigma}+F_{\mu\nu}G_{\rho\sigma}\Bigr)
+\frac{1}{g_4^2}\Bigl(F_{\mu\nu}F^{\mu\nu}+F_{\mu\nu}G^{\mu\nu}\Bigr)\right]\right\}.\cr
&&\label{timerev}
\eea
To construct the 4-dimensional equivalent of Son's fermi-fermi duality, we take the original bosonization duality, 
promote the external gauge field $C_\mu$ to a fluctuating one $\bar{A}_\mu$, (with field strength $\bar{F} = d\bar{A}$), add the terms
\begin{eqnarray}
   \frac{1}{2}\bar{F}\bar{F}-\b \bar{F}G+\a\epsilon \bar{F}G
   +\frac{\gamma}{2}\epsilon \bar{F}\bar{F}\;,
\end{eqnarray}
with constants $\alpha,\beta,\gamma$ that will be fixed in due course, and then integrate out the $\bar A_\mu$. We find that
\begin{eqnarray}
&&\int\!\! {\mathscr D}\bar A_\mu\, {\mathscr D} a_\mu \left\{Z_{\rm fermion}[\bar A;m] \times
\exp \left[+\frac{1}{32\pi} \epsilon^{\mu\nu\rho\sigma} \Bigl(f_{\mu\nu}f_{\rho\sigma} -f_{\mu\nu}\bar{F}_{\rho\sigma}\right.\right.\cr
&&\left.\left.+\a \bar{F}_{\mu\nu}G_{\rho\sigma}+\frac{\gamma}{2}\bar{F}_{\mu\nu}\bar{F}_{\rho\sigma}\Bigr)
+\frac{1}{g_4^2}\left(\frac{1}{2}\bar{F}_{\mu\nu}\bar{F}^{\mu\nu}-\b \bar{F}_{\mu\nu}G^{\mu\nu}\right)\right]\right\}\nonumber\\
&=&\int{\mathscr D}\bar A_\mu\, {\mathscr D}A_\mu\, {\mathscr D}\theta\,\exp\left\{i\!\!\int d^4 x \left[-\frac{(\d_\mu\phi_0)^2}{2}
-\frac{\phi_0^2}{2}(\d_\mu \theta+A_\mu)^2\right.\right.\nonumber\\
&&\left.\left. +\frac{1}{32\pi}\epsilon^{\mu\nu\rho\sigma}\left[F_{\mu\nu}(A)F_{\rho\sigma}(A)+F_{\mu\nu}(A)F_{\rho\sigma}(\bar A)
+\a \bar{F}_{\mu}\bar{F}_{\rho\sigma}+\frac{\gamma}{2}\bar{F}_{\mu\nu}\bar{F}_{\rho\sigma}\right.]
\right.\right.\nonumber\\
&&\left.\left.
+\frac{1}{g_4^2}[F_{\mu\nu}F^{\mu\nu} + F_{\mu\nu}\bar{F}^{\mu\nu} + \frac{1}{2}\bar{F}_{\mu\nu}
\bar{F}^{\mu\nu} -\b \bar{F}_{\mu\nu}G^{\mu\nu}]\right]\right\}.
\end{eqnarray}
Integrating over the $\bar A_\mu$ amounts to using its equation of motion, since it occurs only quadratically in 
the integral, so $(1+\gamma \epsilon)\bar{F} = -F + \b G - \epsilon F -\a \epsilon G$, or
\begin{eqnarray}
  \bar{F}_{\mu\nu} \simeq -F_{\mu\nu}+\frac{1}{2}G_{\mu\nu}
  -\frac{g^2}{32\pi}\epsilon_{\mu\nu\rho\sigma}\Bigl(F^{\rho\sigma}(1-\gamma)+(\a+\b \gamma) 
  G^{\rho\sigma}\Bigr)\,,
\end{eqnarray}
ignoring higher order terms. Substituting this back into the path integral, we obtain on the right-hand side
\begin{eqnarray}
&&\!\!\int\!\! {\mathscr D}A_\mu\, {\mathscr D}\theta\,\exp\left\{i\!\!\int\!\! d^4 x \left[-\frac{(\d_\mu\phi_0)^2}{2}
-\frac{\phi_0^2}{2}(\d_\mu \theta+A_\mu)^2\right.\right.\cr
&&\left.\left. +\frac{1}{32\pi}\epsilon^{\mu\nu\rho\sigma}\left[\frac{\gamma}{2}F_{\mu\nu}F_{\mu\nu} + 
(\b-\a)F_{\mu\nu}G_{\rho\sigma} + \left(\frac{\gamma}{2}\b^2+\a\b \right) G_{\mu\nu}G_{\rho\sigma}\right]\right.\right.\nonumber\\
&&\left.\left.+\frac{1}{2g_4^2}[F_{\mu\nu}F^{\mu\nu} + 2\b F_{\mu\nu}G^{\mu\nu} - \b^2
G_{\mu\nu}G^{\mu\nu}]\right]\right\}.
\end{eqnarray}
We see now that in order to obtain the correct $F^{2}+FG$ contribution, we need $\b=1/2$ (up to a 
rescaling of $g_{4}$) and by comparing to the time-reversed bosonization relation, we also require that $\gamma=-2$ and $\a=3/2$.
Finally, redefining $2g_4^2\equiv \tilde g_4^2$ and using the time-reversed relation (\ref{timerev}), we find
\bea
&&\!\!\int\!\! {\mathscr D}\bar A_\mu\, {\mathscr D} a_\mu\, \left\{Z_{\rm fermion}[\bar A;m] \times
\exp \left[+\frac{1}{32\pi} \epsilon^{\mu\nu\rho\sigma} (f_{\mu\nu}f_{\rho\sigma} - f_{\mu\nu}\bar{F}_{\rho\sigma}\right.\right.\cr
&&\left.\left.+\frac{3}{2} \bar{F}_{\mu\nu}G_{\rho\sigma} - \bar{F}_{\mu\nu}\bar{F}_{\rho\sigma})
+\frac{1}{\tilde g_4^2}\left(\bar{F}_{\mu\nu}\bar{F}^{\mu\nu} - \bar{F}_{\mu\nu}G^{\mu\nu}\right)\right]\right\}\cr
&=&
\!\!\int\!\! {\mathscr D} a_\mu\, \left\{Z_{\rm fermion}[C;m] \times
\exp \left[-\frac{1}{32\pi} \epsilon^{\mu\nu\rho\sigma} \left(f_{\mu\nu}f_{\rho\sigma} - f_{\mu\nu}G_{\rho\sigma}\right.\right.\right.\cr
&&\left.\left.\left.+\frac{1}{2}G_{\mu\nu}G_{\rho\sigma}\right)-\frac{1}{4\tilde g_4^2}G_{\mu\nu}G^{\mu\nu}\right]\right\}.
\eea
It is this then that is the 4-dimensional equivalent of the 3-dimensional fermi-fermi duality of Son, 
and that will fulfil the analogous role in the 4 dimensional duality web. 

\subsection{Bose-bose duality}

Following the strategy in \cite{Karch:2016sxi}, we need to find another version of the bosonization duality that can be 
used twice to obtain a bose-bose duality. To do so, we need to modify the basic bosonization step in order to find just 
the scalar partition function on one side. Continuing as above, we rename the external field $C_{\mu}\to\bar A_\mu$, then add 
\be
\frac{1}{32\pi}\epsilon^{\mu\nu\rho\sigma} G_{\mu\nu} \bar{F}_{\rho\sigma} + \frac{1}{g_4^2}G_{\mu\nu} \bar{F}^{\mu\nu}
\ee
to the action in the bosonization step, then path integrate over $\bar A$. On the scalar side
\begin{eqnarray}
&&\!\!\int\!\! {\mathscr D}\bar A_\mu\, {\mathscr D}A_\mu\, {\mathscr D}\theta\,\exp\left\{i\!\!\int\!\! d^4 x \left[-\frac{(\d_\mu\phi_0)^2}{2}
-\frac{\phi_0^2}{2}(\d_\mu \theta+A_\mu)^2\right.\right.\cr
&&\left.\left. +\frac{1}{32\pi}\epsilon^{\mu\nu\rho\sigma}\Bigl(F_{\mu\nu}F_{\rho\sigma}+F_{\mu\nu}G_{\rho\sigma}
+G_{\mu\nu}\bar{F}_{\rho\sigma}\Bigr)
+\frac{1}{g_4^2}\Bigl(F_{\mu\nu}F^{\mu\nu}+F_{\mu\nu}G^{\mu\nu} + G_{\mu\nu} \bar{F}^{\mu\nu}\Bigr)\right]\right\}\cr
&&
\end{eqnarray}
As before, $\bar A_\mu$ can be integrated out by solving for its equation of motion, which requires that 
$F_{\mu\nu}=-G_{\mu\nu}$. On substituting, and doing the integral over $A_\mu$ (against the delta
 function given by the integration of the Lagrange multiplier), we get
\begin{eqnarray}
&&\!\!\int\!\! {\mathscr D}\theta\, \exp\left\{i\!\!\int\!\! d^4 x \left[-\frac{(\d_\mu\Phi_0)^2}{2}
-\frac{\Phi_0^2}{2}(\d_\mu \theta-S_\mu)^2 + \frac{1}{32\pi}\epsilon^{\mu\nu\rho\sigma}G_{\mu\nu}G_{\rho\sigma} 
+\frac{1}{g_4^2}G_{\mu\nu}G^{\mu\nu}\right]\right\}\cr
&=&Z_{\rm scalar}(-C)\exp\left\{i\int d^4x \left[
+\frac{1}{32\pi}\epsilon^{\mu\nu\rho\sigma}G_{\mu\nu}G_{\rho\sigma}
+\frac{1}{g_4^2}G_{\mu\nu} G^{\mu\nu}\right]\right\}\,.
\end{eqnarray}
Finally, adding in the requisite extra terms on the fermion side, we obtain 
\bea
&&Z_{\rm scalar}(-C) =\!\!\int\!\! {\mathscr D}\bar A_\mu \, {\mathscr D}a_\mu\,  Z_{\rm fermion}(\bar A,m)\,\exp\left\{i\!\!\int\!\! d^4x \left[
+\frac{1}{32\pi}\epsilon^{\mu\nu\rho\sigma}\times\right.\right.\cr
&&\left.\left.\times \Bigl(f_{\mu\nu}f_{\rho\sigma} - f_{\mu\nu}G_{\rho\sigma}
+\bar{F}_{\mu\nu}G_{\rho\sigma} - G_{\mu\nu}G_{\rho\sigma}\Bigr) + \frac{1}{g_4^2}\Bigl(\bar{F}_{\mu\nu}G^{\mu\nu} 
- G_{\mu\nu}G^{\mu\nu}\Bigr)\right]\right\}.\cr
&&
\label{step2}
\eea
At this point we would proceed by using the above bosonization step:  promoting the external field $C$ to a 
dynamical field, say,  $b$; adding in a new external field $A$, as well as the appropriate terms in the exponent and 
integrating over $b$. We would then expect to find the same fermion side of the bosonization step, except in 
terms of the external field $A$ and a time-reversed version, equal to another scalar partition function. Carrying 
out these steps results, however, not in the time-reversed version, but the original partition function! This is so 
because we cannot change the 
sign of the $\epsilon ff$ term, as would be needed for a time-reversed relation. To circumvent this problem, we 
will try to use the {\em same} bosonization step, only now in reverse instead of the time-reversed version.  
Adding only the $1/g_4^2$ terms (with $k_{\mu\nu}\equiv \partial_{[\mu}b_{\nu]}$)
\begin{eqnarray}
   \frac{1}{g_4^2}\Bigl(\beta k_{\mu\nu}F^{\mu\nu}+\gamma k_{\mu\nu}k^{\mu\nu} 
   - \mu F_{\mu\nu}F^{\mu\nu}\Bigr)
\end{eqnarray}
to the action yields a fermion side of the equality that takes the form
\begin{eqnarray}
&&\!\!\int\!\! {\mathscr D}b_\mu\, {\mathscr D}\bar A_\mu\, {\mathscr D}a_\mu\, Z_{\rm fermion}[\bar A,m]\,\exp\Bigl\{i\!\!\int\!\! d^4x \Bigl[
+\frac{1}{32\pi}\epsilon^{\mu\nu\rho\sigma} \Bigl(f_{\mu\nu}f_{\rho\sigma} - f_{\mu\nu}G_{\rho\sigma}
+\bar{F}_{\mu\nu}G_{\rho\sigma}\cr
&-& G_{\mu\nu}G_{\rho\sigma}\Bigr) + \frac{1}{g_4^2}[\bar{F}_{\mu\nu}G^{\mu\nu} - G_{\mu\nu}G^{\mu\nu}
+\beta k_{\mu\nu}F^{\mu\nu} +\gamma k_{\mu\nu}k^{\mu\nu} - \mu F_{\mu\nu}F^{\mu\nu}]\Bigr]\Bigr\}.
\end{eqnarray}
The equation of motion for $b_\mu$ on this side is 
\begin{eqnarray}
2(1-\gamma)k_{\mu\nu}=\bar{F}_{\mu\nu} + \beta F_{\mu\nu}
+\frac{g_4^2}{32\pi}\epsilon_{\mu\nu\rho\sigma}\Bigl(\bar{F}^{\rho\sigma}-f^{\rho\sigma}\Bigr).
\end{eqnarray}
Substituting this into the fermion side of the equality gives an exponent
\bea
&&\frac{1}{g_4^2}\left[\frac{\b}{2(1-\gamma)}\bar{F}_{\mu\nu}F^{\mu\nu}
+\frac{1}{4(1-\gamma)}\bar{F}_{\mu\nu}\bar{F}^{\mu\nu} + \left(\frac{\b^2}{4(1-\gamma)}-\mu \right)F_{\mu\nu}F^{\mu\nu}\right] \cr
&&+\frac{1}{32\pi}\epsilon^{\mu\nu\rho\sigma}\left[\frac{1}{2(1-\gamma)}\bar{F}_{\mu\nu}\bar{F}_{\rho\sigma}
+\frac{\b}{2(1-\gamma)}\bar{F}_{\mu\nu}F_{\rho\sigma} + f_{\mu\nu}f_{\rho\sigma}\right.\cr
&&\left.
+\frac{1}{2(1-\gamma)}f_{\mu\nu}[\bar{F}_{\rho\sigma}+\beta F_{\rho\sigma}]\right].
\eea
Now we note that in order to obtain the correct fermionic side of the bosonization step, we require that
$-2\gamma=2\mu=\b\rightarrow\infty$. Indeed, in this case the exponent in the fermion side of the equality is 
\begin{eqnarray}
&&\frac{1}{g_4^2}\Bigl(F_{\mu\nu}F^{\mu\nu}\Bigr) + \frac{1}{32\pi}\epsilon^{\mu\nu\rho\sigma}
\Bigl(\bar{F}_{\mu\nu}F_{\rho\sigma} + f_{\mu\nu}f_{\rho\sigma} -f_{\mu\nu}F_{\rho\sigma}\Bigr)\;,
\end{eqnarray}
exactly what we have above, with $C$ replaced by $A$, and with the extra terms on the scalar side. 
Finally then, we can put this together as the bose-bose duality, 
\begin{eqnarray}
&&\!\!\int\!\! {\mathscr D} C_\mu\, Z_{\rm scalar}[C]\,\exp\left\{i\!\!\int\!\! d^4x \left[\frac{\b}{g_4^2}
\Bigl(G_{\mu\nu}F^{\mu\nu} -\frac{1}{2}G_{\mu\nu}G^{\mu\nu} - \frac{1}{2}F_{\mu\nu}F^{\mu\nu}\Bigr)\right]\right\}\nonumber\\
&=&Z_{\rm scalar}[-A] \exp\left\{i\!\!\int\!\! d^4x \left[+\frac{1}{32\pi}\epsilon^{\mu\nu\rho\sigma}
F_{\mu\nu}F_{\rho\sigma} + \frac{1}{g_4^2}F_{\mu\nu}F^{\mu\nu}\right]\right\}\;.
\end{eqnarray}
At this point, we need to point out, however, that this relation holds only in the admittedly, 
rather singular $\b\rightarrow\infty$ (in addition to the $g_4\rightarrow 0$) limit.

\section{Conclusions}

Even though the sparks were there for many years prior, it was the discovery of the 3-dimensional duality web 
in the summer of 2016 \cite{Seiberg:2016gmd,Karch:2016sxi,Murugan:2016zal} that really re-ignited the area of 
low-energy dualities, both in the high energy community studying the structure of the gauge/gravity correspondence, 
as well a condensed matter community looking to probe the properties of novel new materials like topological 
insulators and superconductors. The resulting surge of activity has precipitated a plethora of new results at both 
low and high energies as well as an acceleration of the blurring of boundaries between the two disciplines\footnote{As an 
example, we point to the SYK model, currently one of the hottest topics in high energy physics because of its 
leading role in our current understanding of the black hole information problem and the fact that each of S, Y and 
K are condensed matter theorists.}. Among these results was a wonderful article by Karch, Tong and Turner 
\cite{Karch:2019lnn} that realised a 2-dimensional web by collecting several well-known dualities and connecting 
the dots with a non-trivial topological invariant; the Arf invariant that plays a similar role in two dimensions as the 
Chern-Simons invariant in three. Inspired by these 2- and 3-dimensional developments, as well as the more recent 
4-dimensional fermi-fermi duality of Palumbo \cite{Palumbo:2019tmg}, in this article we demonstrate the existence 
of a similar 4-dimensional web of low energy, non-supersymmetric abelian dualities. As with the former dualities, the 
4-dimensional web consists of bose-bose and fermi-fermi dualities as well as a 4-dimensional version of bosonization. 
Any of these can be used to seed the full web. In addition, we have shown that, through dimensional reduction, the 
4 dimensional web can be related to the known 3 dimensional one. Since the latter was shown to descend to the identified 
2 dimensional web in \cite{Karch:2019lnn}, our construction adds the final piece to what is really an {\it inter-dimensional web} 
that relates low-dimensional theories in four, three and two spacetime dimensions. As in the previous duality web 
constructions, there remain some outstanding issues. Among these, we list (in no particular order):
\begin{itemize}
   \item As we pointed out above, the 4 dimensional duality web holds in the infrared. This much was to be expected from the 3
   dimensional case. 
   However, in addition to the low-energy limit, in order to realise the 4dimensional bose-bose duality, and consequently the full web also, 
   we needed to take the (singular) $\beta\to\infty$ limit and while the physics of the $g_{4}\to0$ limit is clear, it is unclear 
   to us how to interpret the former. Clarifying this point would be important in understanding the limits of validity of the 4
   dimensional web, 
   as well as how it relates to the existing lower-dimensional duality webs.
   \item Soon after its discovery, a series of articles by Kachru {\it et.al.} \cite{Kachru:2016rui, Kachru:2016aon} argued that 
   3-dimensional bosonization, understood as a seed of the duality web, could be derived from a single starting point: the 
   3 dimensional mirror symmetry that equates a free chiral superfield and $\mathcal{N}=2$ supersymmetric QED$_{3}$ 
   with a single charged superfield. There, it was shown that when the supersymmetry was broken in a controlled way through a 
   D-term deformation, the chiral duality flows precisely to  the 3 dimensional bosonization duality
   \begin{eqnarray}
      i\bar{\Psi}D\!\!\!\!/_{A}\Psi - \frac{1}{8\pi}AdA\quad \longleftrightarrow \quad |D_{a}\phi|^{2} 
      + \frac{1}{4\pi}ada + \frac{1}{2\pi}Ada\,,
   \end{eqnarray}
   in the infrared. From here, additional 3D dualities can be generated through the action of the modular group of 
   transformations, $(\mathsf{S},\mathsf{T})$, which act on the Lagrangian of a general CFT as 
   \begin{eqnarray}
      \mathsf{S}: \mathscr{L}(\Phi,A) &\mapsto& \mathscr{L}(\Phi,a) - \frac{1}{2\pi}Bda\\
      \mathsf{T}: \mathscr{L}(\Phi,A) &\mapsto& \mathscr{L}(\Phi,A) + \frac{1}{4\pi}AdA\,.
   \end{eqnarray} 
   While the methods utilised in \cite{Kachru:2016rui, Kachru:2016aon} are essentially 3-dimensional, it is interesting
   to note that the (2+1)-dimensional superspace can be obtained from an $\mathcal{N} = 1$ superspace in (3+1) dimensions, 
   by dimensional reduction. The resulting N = 2 superspace has the two basic superfields: 
   A chiral superfield $(\phi, \psi, F)$, where $\phi$ is a complex scalar, $\psi$ a 2-component Dirac fermion and $F$ is 
   an auxiliary complex field, and a vector superfield made up of a gauge field $A_{\mu}$, a real scalar $\sigma$, a 
   2-component Dirac fermion $\lambda$, and an auxiliary real scalar $D$. Given this, and our extensive knowledge of dualities 
   in 4-dimensional supersymmetric theories \cite{Seiberg:1994pq}, it begs the question as to whether a similar unified 
   description of the 4-dimensional duality web exists?
   \item In addition to fuelling some impressive advances in the understanding of 
   Chern-Simons matter \cite{Choudhury:2018iwf, Aharony:2018pjn, Dey:2018ykx}, the 3-dimensional duality web 
   furnishes a powerful non-perturbative framework within which to study a spectrum of strongly correlated quantum 
   critical points. The obvious example here is Son's explanation \cite{Son:2015xqa} of the particle-hole 
    symmetry exhibited by fractional quantum Hall (FQH) states at $\nu=\frac{1}{2}$. More generally, 
   since dual quantum critical points necessarily share a phase diagram, dualities can be used to diagnose the presence of 
   exotic gapped phases of quantum matter. Two more examples that come to mind here are the construction of 
   non-abelian Read-Rezayi FQH states given in \cite{Goldman:2019wvz} and the study of so-called deconfined quantum
   critical points \cite{Wang:2017txt}, exhibited by (2+1)-dimensional quantum magnets such as the non-compact 
   $\mathbb{CP}^{1}$ sigma model coupled to a dynamical non-compact $U(1)$ gauge field $a$ through the covariant derivative 
   $D_{a} = d - ia$,
   \begin{eqnarray}
      \mathscr{L}_{\mathbb{CP}^{1}} = (D_{a}\boldsymbol{z})^{\dagger}(D_{a}\boldsymbol{z}) - 
      \boldsymbol{z}^{\dagger}\boldsymbol{z}\,.
   \end{eqnarray}
   Key insights into such field theories were gained by realizing them on the boundary of  (3+1)-dimensional bulk symmetry 
   protected topological phases, with their 3 dimensional dualities descending from the 4 dimensional 
   bulk theory. In this case - as in the examples
   presented in section 6 of \cite{Seiberg:2016gmd} - the bulk theory plays an auxiliary role, regulating the boundary 
   theory is in a way that preserves the full internal symmetries of the IR fixed point. On the other hand, the IR dualities that are the 
   subject of this article are genuinely 4-dimensional. Can they (or their non-abelian extensions) then be used to predict novel exotic
   gapped phases in (3+1) dimensions? Relatedly, boundaries remain an interesting issue for bosonization 
   \cite{Fukusumi:2021zme, Ebisu:2021acm} and the duality web more generally \cite{Aitken:2017nfd}.  In this case, it would 
   be of interest to know if the class of 4-dimensional theories we consider here can be used to construct phenomenologically
   relevant, anomaly-free theories in (2+1)-dimensions?
\end{itemize}

\section*{Acknowledgements}

The work of HN is supported in part by CNPq grant 301491/2019-4 and FAPESP grants 2019/21281-4 
and 2019/13231-7. HN would also like to thank the ICTP-SAIFR for their support through FAPESP grant 2016/01343-7. 
J.M. was supported in part by the NRF of South Africa under grant CSUR 114599 and in part by a Simons Associateship at the ICTP, Trieste.

\bibliography{4ddualityweb.bib}
\bibliographystyle{utphys}

\end{document}